\documentclass[%
preprint,
 amsmath,amssymb,
 aps,
]{revtex4-1}

\usepackage{graphicx}
\usepackage{color}
\usepackage{dcolumn}
\usepackage{bm}
\usepackage{natbib}
\usepackage{footnote}

\begin{document}

\title{
Financial Brownian particle in the layered order book fluid and Fluctuation-Dissipation relations
}
\author{Yoshihiro Yura$^{1,4}$\footnote{yura@smp.dis.titech.ac.jp}, Hideki Takayasu$^{2,3}$\footnote{takayasu@csl.sony.co.jp}, Didier Sornette$^{4}$\footnote{dsornette@ethz.ch} and Misako Takayasu$^{1}$\footnote{takayasu@dis.titech.ac.jp}}

\affiliation{
$^{1}$Department of Computational Intelligence and Systems Science, Interdisciplinary Graduate School of Science and Engineering, Tokyo Institute of Technology 4259 Nagatsuta-cho, Yokohama 226-8502, Japan.
}
\affiliation{
$^{2}$Sony Computer Science Laboratories, 3-14-13, Higashi-Gotanda, Shinagawa-ku, Tokyo, 141-0022, Japan.
}
\affiliation{
$^{3}$Meiji Institute for Advanced Study of Mathematical Sciences, Meiji University, 
4-21-1 Nakano, Nakano-ku, Tokyo, 164-8525, Japan.
}
\affiliation{
$^{4}$ETH Zurich, D-MTEC, Scheuchzerstrasse 7, 8092 Zurich, Switzerland.
}
\begin{abstract}
We introduce a novel description of the dynamics of the order book of financial markets as that of an effective colloidal Brownian particle embedded in fluid particles. The analysis of a comprehensive market data enables us to identify all motions of the fluid particles. Correlations between the motions of the Brownian particle and its surrounding fluid particles reflect specific layering interactions; in the inner-layer, the correlation is strong and with short memory while, in the outer-layer, it is weaker and with long memory. By interpreting and estimating the contribution from the outer-layer as a drag resistance, we demonstrate the validity of the fluctuation-dissipation relation (FDR) in this non-material Brownian motion process.

\end{abstract}
\maketitle 
The mathematical description of fluctuation phenomena 
in statistical physics, of quantum fluctuation processes 
in elementary particles-fields physics, 
on the one hand, and of financial prices on the other hand, 
has a long tradition since Bachelier's seminal 1900 
PhD thesis\citep{Bachelier1900}, which is anchored in the random walk 
model and Wiener process. In physics, this field of 
study started from Einstein's 1905 paper on Brownian motion, 
the concept of generalized Langevin equation 
and the fluctuation-dissipation relations (FDRs) 
were well-established in the middle of the 
20th century \citep{Kubo,GLE}. Recently, a strong interest 
has developed to clarify the 
conditions under which the FDRs are violated, based on 
numerical simulations~\citep{Sciortino,Kawamura} as well as experimental 
investigations based on precise observations 
using advanced nanotechnology instruments 
for various materials such as colloidal suspension~\citep{Maggi,Strachan}, 
polymers~\citep{Oucris}, liquid crystal~\citep{Joubaud} and 
glass systems~\citep{Komatsu}.

The financial economic literature 
uses the Wiener process as the standard
starting point for modeling and for financial
engineering applications~\citep{Merton}. 
Extending the initial intuition of Bachelier, the random
nature of financial price fluctuations is 
presently mostly understood as resulting
from the the imbalance of buy and sell orders 
at each time step~\citep{Kyle1985}. 
In order to explain non-Gaussian properties of market price fluctuations, 
extensions in the form of Langevin-type equations with an inertia term have been 
proposed~\citep{Hideki,Farmer2002,Bouchaud_Lang,Ide,PRL1997,Takayasu_PUCK,Watanabe,Yura}. 

Essentially all previous models based on the random walk picture or 
its continuous version (the Wiener process) involve just the price dynamics. 
Other approaches simulate financial markets with computational
economic models with different class of agents' strategies 
or using the statistics of buy and sell orders from the viewpoint of statistical physics 
~\citep{PhysRevE.83.016101,Bak1997430,2001cond.mat..1474C,Challet2001285,PhysRevE.64.056136,PhysRevE.71.046131,Takayasu1992127,PhysRevE.79.051120}. 

Here, we introduce a qualitatively novel type of model
for financial price fluctuations. Rather than focusing on the dynamics of a single price for 
a given market that requires complicated modifications to the basic random walk model
in order to account for the numerous stylised facts,
we propose the picture that the observed financial 
motion is analogous to a genuine colloidal Brownian 
particle embedded in a fluid of smaller particles, 
which themselves reflect the structure of the underlying order book 
(defined as the time-stamped list of requests 
for buy and sell orders with prices and volumes). 
The ``Financial Brownian particle in order book molecular fluid'' (in short FBP) picture 
provides a novel quantification of the correlations between different layers in the
order book that can be interpreted as the analogy of the correlation
between a Brownian particle itself with the surrounding fluid molecules. 
We present empirical estimations of the correlation functions that confirm the proposed mapping 
as well as provide non trivial insights on the correlations with deeper fluid molecular layers 
within the order book. 

We analyze the order book data of the Electronic Broking Services 
for currency pairs provided by a market managing company ICAP. 
This foreign exchange market is continuously open 24 hours per day except over weekends 
and its transaction volume per day at about 4 trillion US dollars 
makes it the largest among all financial markets, 
with also much larger liquidity than stock markets. 
We present our results for the US dollar-Japanese Yen market, 
which is characterized by a large transaction volume. 
The traders in this market are international financial companies which 
are connected to the ICAP's market server by a special computer network. 
Any orders, either buy or sell, are quantized by a unit of 1 million US dollar with 
its price given with a granularity of 0.001 Yen (called a pip) 
recorded with a time-stamp of 1 millisecond. 
A pair of buy and sell orders meeting at the same price immediately
triggers a transaction and determines the latest official market price.
These orders disappear from the order book just like 
a pair annihilation of matter-antimatter. 
The price and time quantizations enable us to describe the market 
by particles in discrete space and time, 
where a particle represents either a buy or 
sell order of 1 million US dollar. In the following discussion,
we assign a superscript ``$-$'' (resp. ``$+$'') for 
buy (resp. sell) orders. 

\begin{figure}
\begin{center}
\includegraphics[scale=0.5]{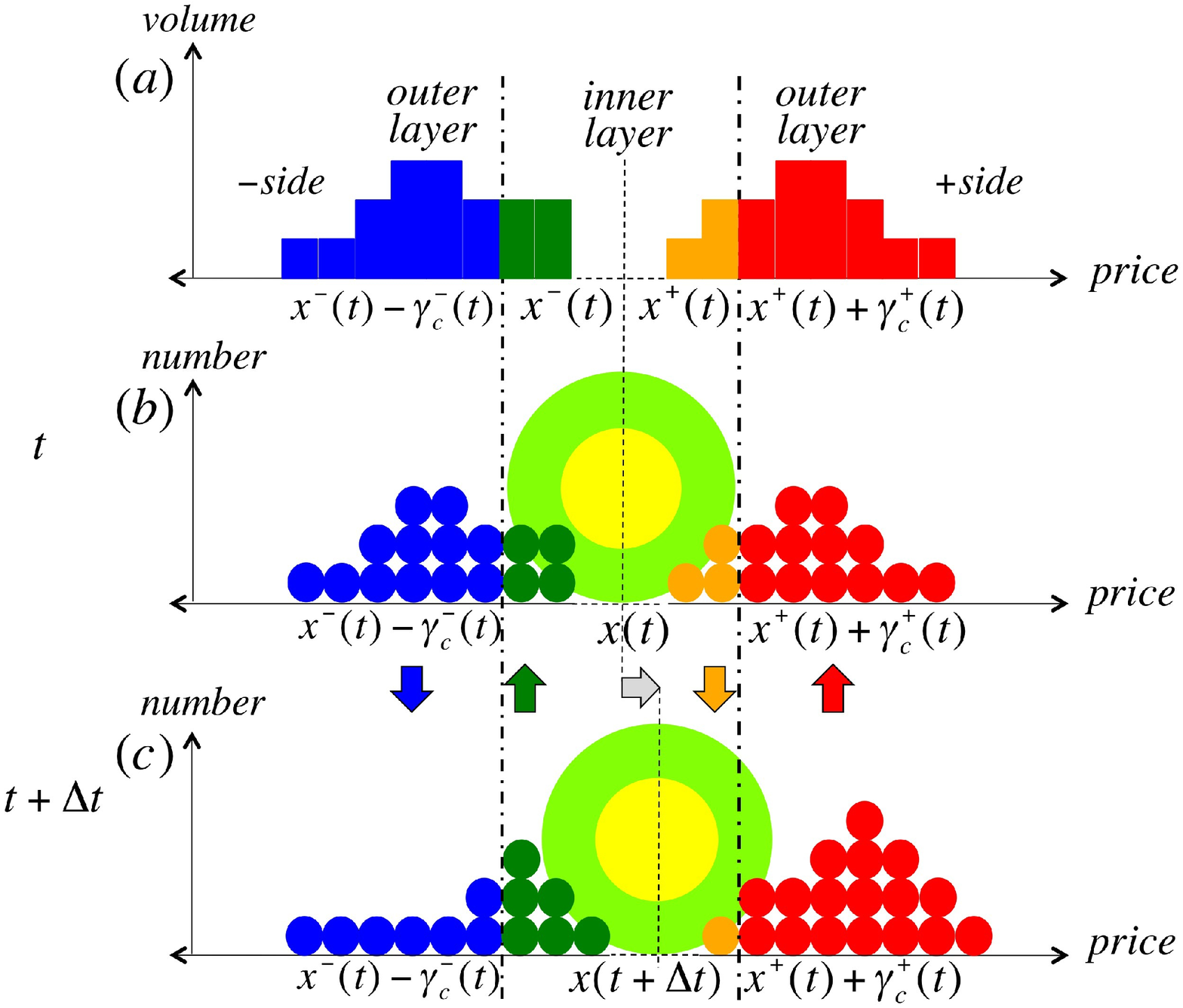}
\end{center}
\caption{
Schematic representation of the FBP model. 
(a) An order book configuration of buy orders (blue and green) 
and sell orders (red and orange) 
on the price axis. 
(b) Corresponding configuration of outer-layer particles (blue and red disks), 
inner-layer particles (green and orange disks), 
the colloidal Brownian particle's interaction range (yellow-green torus), 
and the core (yellow). 
(c) After $\Delta t$, the configuration of 
surrounding particles changes. 
}
\label{fig1}
\end{figure}

At a given time, a state of this market is characterized 
by its order book schematically represented in Fig.~\ref{fig1}(a), 
which contains the set of yet unrealized buy (resp. sell) orders 
in the lower (resp. higher) side of the discrete price axis. 
The highest buy (resp. lowest sell) order price, denoted as $x^{-}(t)$ 
(resp. $x^{+}(t)$) is called the best bid (resp. ask), 
and the gap between the best bid and best ask is called the spread. 
For each buy (resp. sell) order in the order book, we introduce an
important measure of depth $\gamma^{-}$, (resp. $\gamma^{+}$), which is defined by the distance
of this buy (resp. sell) order from $x^{-}(t)$ (resp. $x^{+}(t)$) in pip unit.

The order book is evolved by spontaneous injection of three types of orders; 
limit orders, market orders and cancelation. 
A limit buy (resp. sell) order is introduced by a trader by specifying the buying (resp. selling) price. 
If the buying price is lower than the best ask 
(resp. higher than the best bid) in the order book, 
the order is accumulated at the specified price in the order book as a new buy (resp. sell) order. 
If the buying price is equal to or higher than the best ask (resp. lower than the best bid),
this order makes a deal with a sell order at the best ask (resp. bid) in the order book, 
and those pair of orders annihilate. 
A market buy (resp. sell) order
directly hits against a sell (resp. buy) order at the best
ask (resp. bid) causing a deal. 
A cancelation simply deletes an order, and it can be done only 
by the trader who created the order. 
This highly irreversible particle dynamics 
evokes chemical catalysis, and lead to a rich phenomenology~\citep{Hasbrouck2,Maslov,Bouchaud4,Farmer,Lillo2,Weber,Cont}.

The FBP model that we propose is illustrated in Fig.~\ref{fig1}(b). 
An imaginary colloidal Brownian particle, 
called a colloid, has its center positioned at the mid-price, 
$x(t)= \{x^{-}(t)+x^{+}(t)\}/2$, with the core diameter 
given by the spread, $x^{+}(t)-x^{-}(t)$. 
The accumulated orders are regarded as embedding fluid particles with diameter 
equal to 1 pip. We visualize the core of the colloid by the yellow-disk 
and the interaction range by the yellow-green ring area that
overlaps with the particles near the spread (the green and orange disks). 
We call this interaction range, the inner-layer, 
and the domain outside of this interaction range,  the outer-layer. The values
of threshold depths for defining the inner-layer, $\gamma^{-}_{c}$ and $\gamma^{+}_{c}$, 
will be estimated below from the data. 
With the injection of new orders, the surrounding particles 
change their configuration and the colloid moves as a result, 
as shown in Fig.~\ref{fig1}(c) for a specific example. 
The colored arrows indicate typical 
particle density changes in the layers, which we are going to analyze in detail.

Observing the evolution of the configuration of particles from time $t$ 
to $t+\Delta t$, 
we measure the change in the number of ``$-$'' (resp. ``$+$'') particles as 
a function of the depth $\gamma^{-}$ 
(resp. $\gamma^{+}$), where the depth is measured from $x^{-}(t)$ (resp. $x^{+}(t)$), 
at each time. 
When $x^{-}(t)$ (resp. $x^{+}(t)$) stays at the same location, 
the change of particle number at a given depth is simply given by counting the 
change in the number of
``$-$'' (resp. ``$+$'') particles. When $x^{-}(t)$ (resp. $x^{+}(t)$) moves, 
the density profile as a function of depth shifts accordingly and the changes
of particle numbers at different depth are simply due to the translation.
Note that the depth $\gamma^{-}$(resp. $\gamma^{+}$) can take a negative value, 
for example, in the case when a limit order falls in the spread. 

\begin{figure}
\begin{center}
\includegraphics[scale=0.8]{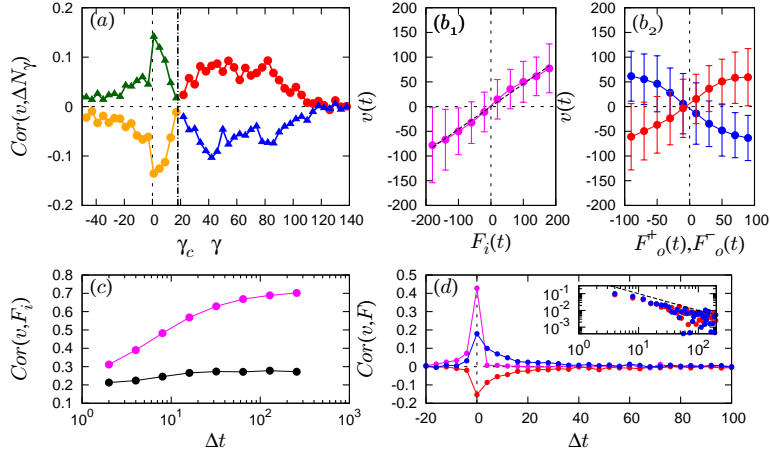}
\end{center}
\caption{
Statistical properties supporting the physical Brownian analogy in FIG.\ref{fig1}. 
(a) Correlation function between velocity, $v(t)$, 
and the change in particle number at depth $\gamma$
for buy orders (green and blue triangles) and sell orders 
(orange and red circles) where $\gamma_{c}=18$. 
(b$_{1}$) Scatter plot of $v(t)$ as a function of $F_i(t)$ for $\Delta t=10^{2}$.
(b$_{2}$) Scatter plot of $v(t)$ as a function of the change in the particle number 
in the outer-layer for buy orders $F_{o}^{-}(t)$ and for sell orders $F_{o}^{+}(t)$ 
with $\Delta t=10^{2}$ (blue line and red line, respectively). 
Dots and bars represent the mean values and the standard deviations. 
(c) Correlation coefficients between
$v(t)$ and $F_i(t)$ (pink dots), and between $v(t)$ and the ``order flow'' (black dots) 
defined by the number of buy orders minus that of sell orders.
(d) Time-shifted correlation functions between
$v(t)$ and \{$F_i(t+\Delta t)$, $F_{o}^{+}(t+\Delta t)$, $F_{o}^{-}(t+\Delta t)$\} 
(pink line, blue line, and red line, respectively). 
The inserted figure show log-log plot for $\Delta t$ 
and the dotted line shows a power law, $1/\Delta t$.
}
\label{fig2}
\end{figure}

Fig.~\ref{fig2}(a) shows the correlations between the change in the number of ``$-$'' 
(resp. ``$+$'') particles at each depth $\gamma$, $\Delta N_\gamma$, 
versus the velocity of the colloid,  $v(t)=(x(t+\Delta t)- x(t))/\Delta t$, where the time is 
in units of tick-time, incremented by one unit when a deal occurs.  
In this figure, the value of the observation window is $\Delta t=100$. 
On average, this time interval corresponds to 160 seconds (1 tick $\simeq$ 1.6 sec). 
The velocity correlation with ``$-$'' particles is negative for 
$\gamma^{-}\le\gamma^{-}_{c}$ as shown by the orange line, 
and the correlation is positive for $\gamma^{-}_{c}<\gamma^{-}$ shown by the red-line,
where the estimated value of the boundary is $\gamma^{-}_{c}=18$.
The same relations with the opposite sign hold 
for ``$+$'' particles, as shown by the green and blue lines, implying that the dynamics is close to symmetric. 
Intuitively, when the price goes up, more new buy than sell orders are injected
in the inner range and sell orders near the market price tend to be canceled 
and replaced by higher sell prices, based on the anticipation 
of larger future returns by traders assuming trend persistence. The opposite direction is explained in the same way.

Let us define the total change of particle numbers in the inner-layer at time $t$, 
$f_{i}(t)=c^{-}_{i}(t)-c^{+}_{i}(t)-a^{-}_{i}(t)+a^{+}_{i}(t)$ 
where $c^{-}_{i}(t)$ and $a^{-}_{i}(t)$ denote the numbers of ``$-$'' 
particles that are created and annihilated, respectively, in the inner-layer at time $t$, 
and  $c^{+}_{i}(t)$ and $a^{+}_{i}(t)$ are the same quantities for ``$+$'' particles. Note that ``$-$'' and ``$+$'' 
particles are counted with the opposite sign as they are conjugate 
``matter'' and ``anti-matter''. 
In Fig.~\ref{fig2}(b$_{1}$), the scatter plot of the velocity of the colloid, $v(t)$,
observed in the same time window, $\Delta t=100$, as a function of 
 the sum  $F_{i}(t)=\sum_{s=0}^{\Delta t}f_{i}(t+s)$ from $t$ to $t+\Delta t$
 demonstrates a strong linear correlation.
Fig.~\ref{fig2}(c) shows the correlation coefficient between $v(t)$ and $F(t)$
for different values of $\Delta t$. The correlation increases for larger $\Delta t$ 
reaching the value $0.7$ around $\Delta t=100$. These empirical results
suggest the following basic relation:
\begin{eqnarray}
v(t)=L(\Delta t)\frac{1}{\Delta t}\sum_{s=0}^{\Delta t}f_{i}(s)+\eta(t) ~.
\label{velocity}
\end{eqnarray}
The factor $L(\Delta t)$ represents the mean step length of the colloid motion 
as a response to the motion of the surrounding fluid
particles. It ranges from $L(\Delta t) \approx 0.44$ pips for $\Delta t=100$ to
$L(\Delta t) \approx 0.32$ pips for $\Delta t=4$. The last term $\eta(t)$ is the error term. 
Similar correlation between the velocity and market orders have been previously
documented \citep{Hasbrouck2}. However, we find here significantly higher values
due to the more appropriate separation of the negative and positive sides of the layered structure
of the order book.
We also observed similarly the changes in outer-layer particle numbers 
for buy orders and sell orders separately, 
and confirmed the correlations with the velocity as shown in Fig.~\ref{fig2}(b$_{2}$). 
Fig.~\ref{fig2}(d) shows the time-shifted correlations between 
the velocity and these changes of particle numbers, which confirms
that the inner-layer particles correlate strongly with the velocity but with a short memory,
while the outer-layer particles' correlation is weaker but decays slowly with a power tail.  

\begin{figure}
\begin{center}
\includegraphics[scale=0.5]{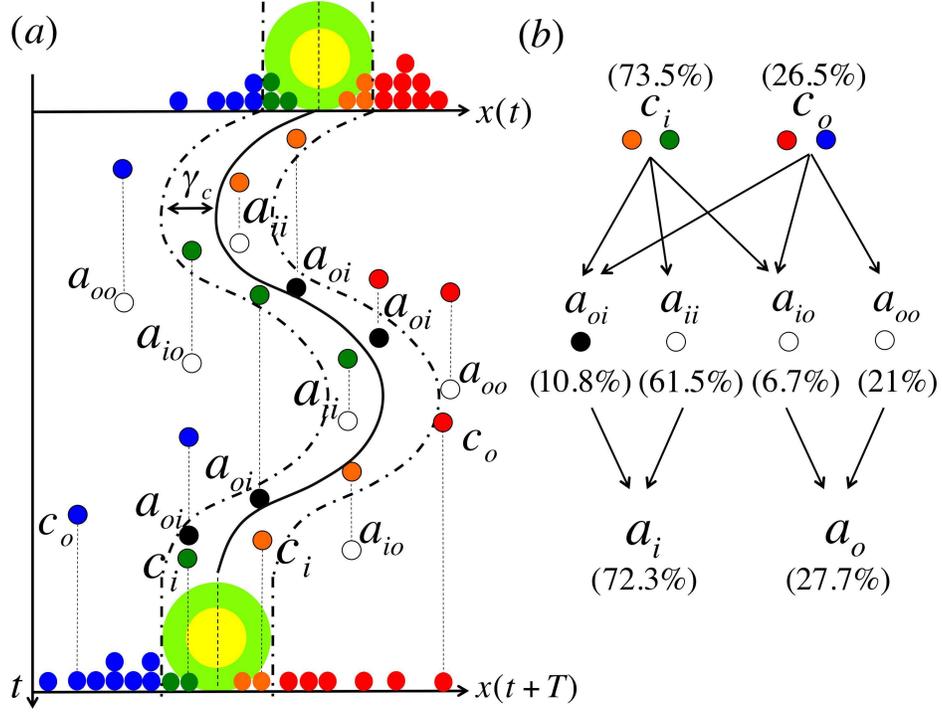}
\end{center}
\caption{
Schematic diagram of the space-time configuration of particles. 
(a) Creation and annihilation of fluid particles 
are shown together with the trajectory of the colloid motion (black line) 
and the inner-layer (the region between black dotted lines). 
The black disks represent $a_{oi}$ particles. 
(b) Share percentages of the different particle types. 
}
\label{fig3}
\end{figure}

Next, we present a more detailed description of the fluid particles. 
Fig.~\ref{fig3}(a) shows a schematic diagram of the space-time configuration. 
We categorize the particles that annihilate in the inner-layer into two classes, $a_{ii}$ and $a_{oi}$. 
An $a_{ii}$ particle is created in the inner-layer, 
stays in the inner layer and annihilates in the inner-layer, while an $a_{oi}$ particle is 
either born in the inner-layer or outer-layer, visits the outer-layer 
at least once and then annihilated in the inner-layer. 
By surveying the whole particle's life, we find that 73.5$\%$ of particles 
are created in the inner-layer (denoted as $c_{i}$), and 72.3$\%$ of particles 
are annihilated in the inner-layer (denoted as $a_{i}$). 
The share of $a_{ii}$ particles is 61.5$\%$ and that of $a_{oi}$ is 10.8$\%$. 

\begin{figure}
\begin{center}
\includegraphics[scale=0.8]{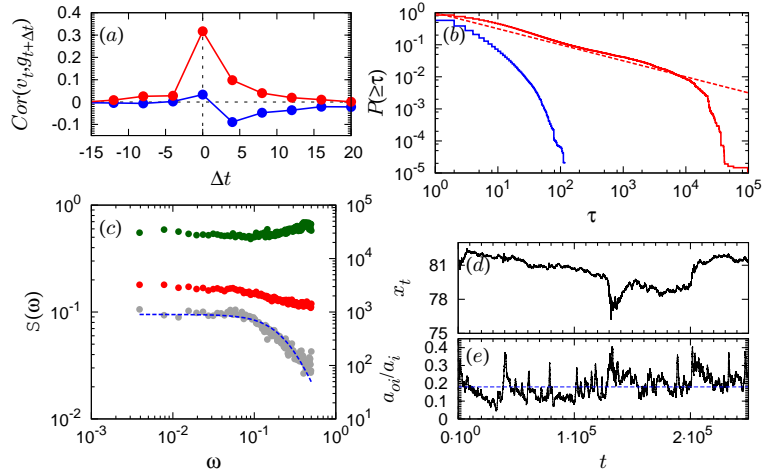}
\end{center}
\caption{
\space
Statistical properties of $a_{oi}$ particles (see text for definition): 
(a) Correlation functions of 
$\{v(t),g_{oi}(t+\Delta t)\}$ (red) and 
$\{v(t),g_{ii}(t+\Delta t)\}$ (blue). 
(b) Cumulative lifetime distribution of $a_{ii}$(blue) 
and $a_{oi}$(red), with the power law $\tau^{-0.5}$ shown as a guideline (red dotted line). 
(c) Power spectra of $v(t)\Delta t$ (green), $v_{I}(t)\Delta t$ (red), and ratio of $|v_{I}^{2}(\omega)|/|v^{2}(\omega)|$ 
(gray), with the blue dotted guideline $a/(b+\omega^{2})$ with
(a,b)=(0.00725, 0.0764). 
The power spectra are averaged over 200 samples of size $2^{8}$ ticks time series. 
The right vertical axis is for $v(t)\Delta t$ and $v_{I}(t)\Delta t$, 
and the left vertical axis is for the ratio. 
(d) Time series of $x(t)$ (USD-JPY exchange rates for the week of March 14, 2011).
(e) Time series of the ratio, $a_{oi}$/$a_{i}$, 
measured for the time interval, [$t-1000$, $t$], 
and the averaged value for the whole period(the blue dotted line).
}
\label{fig4}
\end{figure}

Statistics of $a_{ii}$ and $a_{oi}$ are compared 
in Fig.~\ref{fig4}. 
The time-shifted correlations with $v(t)$ 
are plotted in Fig.~\ref{fig4}(a) for $g_{ii}(t)=-a_{ii}^{-}(t)+a_{ii}^{+}(t)$ 
by blue line and for $g_{oi}(t)=-a_{oi}^{-}(t)+a_{oi}^{+}(t)$ 
by red line. We find that $a_{ii}$ and $a_{oi}$ are oppositely correlated with the velocity. 
Fig.~\ref{fig4}(b) shows the cumulative distributions of lifetime 
of these particles in log-log scale. 
The distribution of the lifetimes of $a_{ii}$ particles decays exponentially with a mean lifetime 
of approximately 2.6 ticks, while that of $a_{oi}$ follows a power law with an exponent 
close to $-0.5$,  which corresponds to the distribution 
of recurrence time intervals for 1-dimensional random walks. 

These results justify a more sophisticated FBP picture in which
the $a_{ii}$ particles contribute to 
the driving force, directly pushing or pulling the colloid at the time of annihilation, 
and the $a_{oi}$ particles work as a drag  
that impede the colloidal motion since they always collide 
with the front of the colloidal motion. Based on this picture, the velocity equation, Eq.(\ref{velocity}), 
is decomposed as $v(t)=v_{I}(t)+v_{F}(t)+\eta(t)$ 
where 
$v_{I}(t)=L(\Delta t)\sum_{s=0}^{\Delta t}\{-a_{oi}^{-}(s)+a_{oi}^{+}(s)\}/\Delta t$, 
$v_{F}(t)=L(\Delta t)\sum_{s=0}^{\Delta t}\{c_{i}^{-}(s)-c_{i}^{+}(s)-a_{ii}^{-}(s)+a_{ii}^{+}(s)\}/\Delta t$. 

As the term $v_{F}(t)$ is nothing but the direct driving force term that
reflects the immediate orders of traders, 
we focus on the term $v_{I}(t)$, which is caused by the long-term response
of the fluid particles.
The power spectra of $v(t)\Delta t$, $v_{I}(t)\Delta t$, and their ratio, 
are plotted in Fig.~\ref{fig4}(c). 
The power spectrum of $v(t)$ 
is nearly white with slightly more energy in the high frequency band,
implying that there are zig-zag fluctuations at very short times. 
On the other hand, the spectrum of $v_{I}(t)\Delta t$ clearly decays at high frequency. 
The ratio of power spectra,
$|v_{I}(\omega)|^{2}/|v(\omega)|^{2}$, 
has a Lorentzian form, implying 
that the response function is approximated by an exponential function, 
$v_{I}(t)=\int_{-\infty}^{t}\phi(t-s)v(s)ds$, 
$\phi(t)=\phi_{0}e^{-\delta t}$ 
with the following estimated parameter values, $\phi_{0}=0.08$ and $\delta=0.27$. 
Introducing these relations into $v(t)=v_{I}(t)+v_{F}(t)+\eta(t)$ 
and taking its time derivative, 
we obtain the following standard form of a Langevin equation in the continuous time representation, which demonstrates the validity of the FDR: 
\begin{eqnarray}
\frac{d}{dt}v(t)=-\mu v(t)+G(t),~~~~ \mu=\frac{\delta }{\phi_{0}}-1 ~.
\end{eqnarray}
The external force term $G(t)$ 
includes $v_{F}$, $\eta$ and their derivatives, 
which are not simple white noises, and the drag coefficient 
is estimated as $\mu=2.2$. The validity of this continuum formulation
can be checked by estimating the Knudsen number \citep{Knudsen,Fluid,PhysRevE.74.031402} of the financial markets, 
defined as the ratio of the mean free path of collisions of the colloid and the $a_{oi}$ particles 
over the diameter of the colloid. We find that the averaged value of the Knudsen number is
approximately $0.02$, whose smallness  
guarantees the validity of the continuum representation of market price given by Eq. (2).

So far, we have analyzed the whole data set to observe the averaged behaviors, 
neglecting the well-known fact that markets are not stationary and are characterized
by regime shifts: calm periods are punctuated 
by turbulent periods of high transient volatility including speculative bubbles and crashes. 
It is thus more appropriate to revisit our above estimations of observables
in shorter time-scales and analyze their possible time variation.
A detailed analysis will be described in a separate paper. Here,
we show the temporal change of the ratio of $a_{oi}/a_{i}$ 
in Fig.~\ref{fig4}(e) with the corresponding market price 
in Fig.~\ref{fig4}(d). In addition to being easy to observe, the ratio $a_{oi}/a_{i}$ 
constitutes the key parameter related to the strength of the drag force exerted by the fluid particles. 
One can see that  $a_{oi}/a_{i}$ fluctuates significantly, confirming
that market conditions are not stationary.

In summary, we have established a fundamental analogy between 
the motion of a colloidal particle embedded
in a fluid and the price dynamics of a financial market in the order book. 
By observing the detailed behaviors of the colloid and surrounding particles in the order book, 
we found that the drag resistance is caused by particles moving from the outer-layer to
the inner-layer. The proposed quantitative correspondence provides
a novel perspective for the analysis of financial markets.
In addition, it should provide a
stimulus for physicists from many different fields, since
the question of the origin of drag resistance is
a fundamental question in physics that still remains to be fully clarified.
We also showed the need to enhance the analysis by accounting for
the non-stationary properties of markets. Moreover, there are some market conditions
when we find that Knudsen number becomes close to $1$, requiring to 
extend the continuous description (2) into  a discrete representation. 
Our approach demonstrates the importance of physical intuition
associated with financial insights to analyze big data of financial markets. 

The authors appreciate useful comments by Prof. Hisao Hayakawa. 
This work was partially supported by Grant-in-Aid for Scientific Research (C) 24540395. 
Y.Y. was financially supported by Japan Society for the Promotion of Science, 
Research Fellowship for Young Scientists.

\bibliographystyle{apsrev4-1}
\bibliography{prl3} 

\end{document}